\documentstyle[12pt]{article}
\def\beqar {\begin{eqnarray}}
\def\eeqar {\end{eqnarray}}
\def\beq {\begin{equation}}
\def\eeq {\end{equation}}
\def \ep {{\epsilon}}
\def \half {{\textstyle{1\over 2}}}
\def \l {{\lambda}}
\def \la {{\langle}}
\def \ra {{\rangle}}

\def \Tr {{\rm Tr}}
\def \bz {{\bar z}}
\begin{document}

\begin{titlepage}
\null\vspace{-62pt}

\pagestyle{empty}
\begin{center}
\rightline{}
\rightline{CCNY-HEP-00/3}
\rightline{RU-00-8-B}

\vspace{1.0truein} {\large\bf Quantum Mechanics on a
Noncommutative Brane}\\
\vskip .1in {\large\bf in M(atrix) Theory}

\vspace{1in} V. P. NAIR\\
\vskip .1in {\it Physics Department\\ City College of the CUNY\\
New York, NY 10031}\\
\vskip .05in and
\vskip .05in {\it Physics Department\\ Rockefeller University\\
New York, NY 10021}\\
\vskip .05in {\rm E-mail: vpn@ajanta.sci.ccny.cuny.edu}\\
\vspace{1.5in}

\centerline{\bf Abstract}

\end{center}
\baselineskip=18pt

We consider the quantum mechanics of a particle on a
noncommutative two-sphere with the coordinates obeying an
$SU(2)$-algebra. The momentum operator can be constructed in
terms of an
$SU(2)\times SU(2)$-extension and the Heisenberg algebra
recovered in the smooth manifold limit. Similar considerations
apply to the more general $SU(n)$ case.
\end{titlepage}

\hoffset=0in
\newpage
\pagestyle{plain}
\setcounter{page}{2}
\newpage

question of setting up the quantum mechanics of a particle on a
brane configuration in the matrix model of $M$-theory \cite{bfss,
taylor}. It is by now clear that the matrix model can
successfully describe many of the expected features of
$M$-theory. Smooth brane configurations and solutions of
$M$-theory can be obtained in the large
$N$-limit of appropriate $(N\times N)$-matrix configurations
\cite{taylor, branes}. Now, brane solutions in  M(atrix) theory
are examples of noncommutative manifolds, specifically those with
an underlying Lie algebra structure. The relationship between the
matrix description and $M$-theory and strings suggests that
noncommutative manifolds with an underlying Lie algebra
structure  (or their specializations into cosets) would be the
most interesting ones from a physical point of view. Therefore we
shall focus on such manifolds, although one can, of course,
consider the question of quantum mechanics on more general
noncommutative manifolds as well.

There is, by now, an enormous number of papers dealing with
noncommutative geometry. One line of development has to do with
spectral actions and the use of the Dirac operator to
characterize the manifold, motivated by quantum gravity
\cite{connes}. Quantization of such actions has also been
attempted
\cite{rovelli}. The majority of recent papers deals with
noncommutative manifolds with an underlying canonical structure
and the construction of field theories on these spaces
\cite{witten}. There has also been some recent work on manifolds
with an underlying Lie algebra structure, including the
definition of a star product and the construction of gauge fields
which take values in the enveloping algebra \cite{wess}. The
topic of the present paper fits within the general milieu of
these ideas and investigations, but we also have a very specific
theoretical context, namely, brane solutions in M(atrix) theory.
If the world is a brane \cite{randall}, and if it is realizable
as a solution in M(atrix) theory, then the quantum mechanics of a
particle on a brane is clearly of more than mathematical interest.

Consider a particular brane solution in M-theory, say, the
noncommutative spherical membrane. In this case the brane has the
topology of $S^2\times S^1$, where the $S^2$ is the
noncommutative part described by matrices and $S^1$ denotes the
compactified 11-th dimension. The two-sphere is given in terms of
three matrix coordinates which may be taken as
\beq Q_a = { r\over \sqrt{j(j+1)}} ~t_a \label{1}
\eeq where $r$ is a fixed number which is the radius of the
sphere and $t_a,~ a=1,2,3$, are the generators of $SU(2)$ in the
$(2j+1)$-dimensional matrix representation. As the dimension
$(2j+1) \rightarrow \infty$, we get a smooth manifold which is
$S^2$. This limit can be very explicitly understood by the
representation
\beqar &&T_+ = T_1 +i T_2 = z^2 \partial_z -N z\nonumber\\ &&T_-
=T_1 -iT_2 = -\partial_z \nonumber\\ &&T_3 = z \partial_z -\half
N \label{2}
\eeqar A basis of states on which these act is given by $\vert
\alpha \ra ,~\alpha = 0,1,...,N$ with
$\la z\vert \alpha \ra = 1, z, z^2,..., z^N$. The inner product
is given by
\beq
\la g \vert f \ra = (N+1) \int {d^2z\over \pi} {1\over (1+z \bz
)^{N+2}} ~{\bar g} ~f
\label{3}
\eeq The matrix elements of $T_a$ with the $(N+1)$ states $\vert
\alpha \ra$ give the standard matrix version of $t_a$, {\it
viz.}, $(t_a)_{\alpha \beta} =
\la \alpha \vert T_a \vert \beta \ra$. By partial integration, we
can see that
$T_a$ can be replaced, in matrix elements, by
\beqar &&T_+ = \l \phi +z^2\partial_z \nonumber\\ &&T_- = \l 
{\bar \phi} -\partial_z \nonumber\\ &&T_3 = \l \phi_3 + z
\partial_z \label{4}
\eeqar where $\l=\half (N+2)$ and
\beq
\phi = {2z\over 1+z\bz },~~~~~~~~~~~~{\bar \phi}= {2\bz \over
1+z\bz},~~~~~~~~~~~~
\phi_3 = {1-z\bz \over 1+z\bz}\label{5}
\eeq As $N\rightarrow
\infty$, the
$\l\phi$-terms in the above expressions for $T_a$ dominate and we
find $t_a
\rightarrow \l \phi_a$. Thus
$Q_a\rightarrow r \phi_a$, with $\phi_a\phi_a =1$. The membrane
is described by the continuous coordinates $z,\bz$.

At finite $N$, the two-sphere is described by the $(N+1)$ states
$\vert \alpha
\ra$ which may be thought of as approximating the sphere by
$(N+1)$ points, none of which has sharply defined coordinates.
Translations of $\vert\alpha
\ra$ can be achieved by the use of $T_{\pm}$. However, this is
not what we want. As $N\rightarrow
\infty$, $T_{\pm}$ go over to $\phi ,{\bar \phi}$ and correspond
to the mutually commuting coordinates $z, \bz$. They do not play
the role of momenta conjugate to those coordinates obeying the
Heisenberg algebra. We need to identify the momenta which lead to
the Heisenberg algebra as
$N\rightarrow \infty$. Since the latter does not have finite
dimensional matrix representations, it is also clear that we
should expect a modified algebra at finite $N$. Ultimately, from
the point of view of noncommutative spaces, one keeps $N$ finite,
the limit being taken only to show agreement with the smooth
manifold limit.

The classical dynamics of a particle moving on a sphere gives a
clue to the choice of a momentum operator or generator of
translations. In the classical case, we may write the momentum as
$P_a = (1/q^2) \epsilon_{abc} q_b J_c$, where $q_a$ is the
coordinate and
$J_c$ is the angular momentum operator, here taken as the
fundamentally defined quantity.  (If we reduce $J_c$ in terms of
$q_b$, and $p_b$ conjugate to it, we find $P_a = \left(
\delta_{ab} - q_aq_b/q^2\right)p_b$, which are the correct
translation  generators consistent with $q_aq_a=1$.) Absorbing
$r$ into the definition of $Q_a$, a possible choice of $P_a$ is
then $ -\half (\l /Q^2)
\epsilon_{abc} (Q_bJ_c +J_c Q_b )$, where we have symmetrized
$Q_a, J_a$ to form a hermitian combination. The operators $Q_a,
J_a$ obey the algebra
\beq [Q_a ,Q_b] = {i\over \l} \ep_{abc} Q_c\label{6}
\eeq
\beqar &&[J_a, Q_b] = i \ep_{abc} Q_c \nonumber\\ &&[J_a, J_b] =
i\ep_{abc} J_c 
\label{7}
\eeqar Notice that 
\beqar
\half \ep_{abc} (Q_bJ_c +J_c Q_b) &&= \ep_{abc} \left( Q_bJ_c
+{i\over 2}
\ep_{cbk} Q_k\right)\nonumber\\ &&= \ep_{abc} Q_bJ_c
-iQ_a\nonumber\\ &&=
\ep_{abc} Q_b (J_c -\l  Q_c)\nonumber\\ &&=\l \ep_{abc} Q_b
K_c\label{8}
\eeqar where $\l K_a =J_a -\l Q_a$. Further
\beqar &&[K_a ,K_b ] = {i\over \l}\ep_{abc} K_c \nonumber\\
&&[K_a, Q_b]=0\label{9}
\eeqar Therefore, rather than starting with $J_a$, we might as
well consider the mutually commuting
$SU(2)\times SU(2)$-algebra of $Q_a, K_a$ and define the momentum
operator as
\beq P_a = \l {\ep_{abc} \over \sqrt{Q^2 K^2}} K_b Q_c \label{10}
\eeq where $Q^2 = Q_aQ_a, ~K^2 =K_a K_a$. Obviously, $[P_a, Q^2]
=[P_a, K^2]=0$ so that there is no ordering ambiguity in the
definition of $P_a$. We have chosen to divide by the symmetric
expression
$\sqrt {Q^2K^2}$ eventhough the classical expression had $q^2$.
As we shall see below, $Q^2\approx K^2$ in the continuous
manifold limit. Also the parameter
$\l$ will be related to $Q^2, K^2$ below. The commutation rules
for $P_a$ become
\beqar &&[P_a, Q_b]~= {i\over \sqrt{Q^2K^2}}\left[ \delta_{ab}
Q\cdot K -\half (Q_aK_b+Q_bK_a)  -{\epsilon_{abc}P_c\over
2\l}\right] \nonumber\\ &&[P_a, P_b]~= i\ep_{abc} {Q\cdot K\over
Q^2K^2} ~J_c\label{11}
\eeqar
$J_a =\l (Q_a +K_a)$ are the generators of the diagonal $SU(2)$
subgroup.

The smooth manifold limit can be understood by considering large
representations for $Q_a$ and
$K_a$, and analyzing representations of the diagonal $SU(2)$ of
$J_a$. Labelling the corresponding spins by lower case letters,
we find $\l^2 Q^2 = q(q+1),~\l^2 K^2 =k(k+1),~J^2 =j(j+1)$ and
$2\l^2 Q\cdot K= j(j+1)- q(q+1) -k(k+1)$. If we take $q,k$ very
large and the combined spin $j$ to be small and fixed, and $\l^2
= \sqrt{Q^2K^2}\approx q(q+1)\approx k(k+1)$, we find that the
algebra (\ref{6}), (\ref{9}), (\ref{11}) reduces to
\beqar &&[Q_a,Q_b]~\approx 0\nonumber\\ &&[P_a, Q_b]~\approx
-i\left(
\delta_{ab} - {Q_aQ_b \over Q^2}\right)\label{12}\\ &&[P_a,
P_b]~\approx -i\ep_{abc} {J_c\over Q^2}\nonumber
\eeqar We also have $Q_a Q_a \approx 1$. Equations (\ref{12}) are
the Heisenberg algebra restricted to a smooth two-sphere of unit
radius. (Taking
$\l^2 = \sqrt{Q^2K^2}/r^2$, we can get a radius equal to
$r$.) 

The emergence of the continuous coordinates and the large
$\l$-expansion can be seen in more detail as follows. We write a
general
$SU(2)$-valued $(2\times 2)$-matrix as
\beq g = \sqrt{1-{x^2\over r^2}} ~+~i {\vec\sigma}\cdot {{\vec
x}\over r}
\label{13}
\eeq where $\sigma_a$ are the Pauli matrices. We then find that
\beqar &&g^{-1}~dg = i ~{\sigma_a \over 2}~ E_{ab} dx^b
\nonumber\\ &&dg ~g^{-1} = i ~{\sigma_a\over 2} ~{\tilde E}_{ab}
dx^b\label{14}
\eeqar where
\beqar &&E_{ab} ~= {2\over r^2}\left[ -\ep_{abc} x_c
+{\delta_{ab} (r^2-x^2) +x_ax_b \over
\sqrt{r^2-x^2}}\right]\nonumber\\ &&{\tilde E}_{ab} ~=
E_{ba}\label{15}
\eeqar The above equations define the frame fields on $SU(2)$.
The inverses to
$E_{ab}$,
${\tilde E}_{ab}$ are given by
\beqar &&E^{-1}_{ab} =\half \left( \ep_{abc} x_c +\delta_{ab}
\sqrt{r^2-x^2}\right)\nonumber\\ &&{\tilde E}^{-1}_{ab} =
E^{-1}_{ba}\label{16}
\eeqar The quantities
\beqar &&Q'_a ~= ~~i{\tilde E}^{-1}_{ka} {\partial \over \partial
x_k}\nonumber\\ &&K'_a ~= -i { E}^{-1}_{ka} {\partial \over
\partial x_k}
\label{17}
\eeqar obey mutually commuting $SU(2)$ algebras. Further, since
$[ x_a,
\sqrt{r^2-x^2}\partial_b] -[x_b, \sqrt{r^2-x^2}\partial_a] =0$,
we see that we can shift $Q'_a$ by $x_a$ and $K'_a$ by $-x_a$ and
still obtain the same algebra. In other words, we can define
\beqar &&Q_a ~= \left[ ~~{x_a} +{i\over 2 \l} \left( \ep_{abc}
{x_c} +\delta_{ab} 
\sqrt{r^2-x^2}
\right){\partial \over \partial x_b}\right]\nonumber\\ &&K_a~=
\left[ -{x_a} +{i\over 2 \l} \left( \ep_{abc} {x_c} -\delta_{ab} 
\sqrt{r^2-x^2}
\right){\partial \over \partial x_b}\right]\label{18}\\ &&J_a~=\l
(Q_a +K_a) = -i\ep_{abc} x_b {\partial \over \partial
x_c}\nonumber
\eeqar This is in a form suitable for the large $\l$-expansion
for $SU(2)\times SU(2)$, with the combined total spin being
small. As $\l\rightarrow \infty$, the $x_a$-terms are dominant in
the expressions for $Q_a, ~K_a$ and we get
$Q_a\rightarrow x_a, ~ K_a \rightarrow -x_a$.  The algebra
(\ref{6}), (\ref{9}), (\ref{11}) reduces to
\beqar &&[P_a, Q_b]~\approx -i\left( \delta_{ab} - {x_ax_b \over
x^2}\right)
\nonumber\\ &&[P_a, P_b]~\approx -i \ep_{abc} {J_c \over
x^2}\label{19}
\eeqar
$x^2=x_ax_a =r^2$ is a constant in this limit. The $\phi$'s given
in (\ref{5}) are a particular parametrization of the $x_a$'s
subject to $x_ax_a$ being constant.

In taking the limit as above we have retained $S^2$-topology for
the smooth manifold. It is important to realize that since we are
dealing with $Q$'s which obey a Lie algebra, $Q^2$ is fixed for
any representation and hence we will not get a flat Heisenberg
algebra. A way to obtain the flat space algebra would be to take
the radius $r$ to be very large and then restrict the operators
to a small neighbourhood on the sphere. This will lead to a flat
two-dimensional Heisenberg algebra as $r\rightarrow \infty$.  For
example, we can expand around
$x_a= (0,0,r)$. It is interesting to see how this works out
directly in terms of the operators $Q_a, ~K_a$. The neighbourhood
of $x_a= (0,0,r)$ corresponds to $Q_3$ and $-K_3$ being large.
Since
$Q_3 \sim r$ and $\l \sim k/r$, we see that $[Q_1,Q_2]\sim
ir^2/k$ and so, the commutativity of coordinates in the large
$k$-limit requires that $r^2 \sim k^\delta$ with $ \delta < 1$ as
$k\rightarrow \infty$. On the other hand, we also have
$[P_1,P_2]\sim 1/r^2$ and the vanishing of this requires
$\delta >0$. The simplest and symmetrical choice is to take
$\delta =\half$ or
$r\sim k^{1\over 4}$. We define eigenstates of $Q_3,~K_3$ by
\beqar &&K_3 ~\vert m,n\ra = (-k+m) ~\vert m,n\ra\nonumber\\
&&Q_3 ~\vert m,n\ra = (~~k-n) ~~\vert m,n\ra \label{20}
\eeqar Restricting to small neighbourhood of large $Q_3$, $-K_3$
means that the integers $m,~n$ can be considered to be small
compared to $k$. In this case, introducing raising and lowering
operators
$\alpha^\dagger ,~\alpha$ for $n$ and $\beta^\dagger,~\beta$ for
$m$,  we find
\beqar &&Q_1 = {r\over \sqrt{2k}} ~(\alpha +\alpha^\dagger
)\nonumber\\ &&Q_2 = {ir\over \sqrt{2k}} ~(\alpha^\dagger -\alpha
)\nonumber\\ &&P_1 = -{i\over r}
\sqrt{k\over 2} ~(\alpha -\alpha^\dagger +\beta^\dagger -\beta
)\label{21}\\ &&P_2 = -{1\over r} \sqrt{k\over 2} ~(\alpha
+\alpha^\dagger +\beta^\dagger +\beta )\nonumber
\eeqar where we can take $r=r_0 k^{1\over 4}$ and, as usual,
$\alpha \vert m,n\ra = \sqrt{n}~\vert m,n-1\ra$,
$\beta \vert m,n\ra =\sqrt{m}~ \vert m-1,n\ra$, etc. The flat
space Heisenberg algebra is now easily verified.

In the usual procedure of quantization, starting with a set of
classical coordinates $q_a$, one introduces the momenta and the
phase space, thereby doubling the number of variables. The
quantum theory is then defined by one irreducible representation
of the Heisenberg algebra. equivalent to the standard
Schr\"odinger representation. The restriction to irreducibility
is equivalent to the requirement that the wavefunctions depend
only on half of the phase space variables, the coordinates
$q_a$, for example. This is the so-called polarization condition.
For a given wavefunction, this allows the determination of the
momenta as $p_a\psi = -i({\partial \psi / \partial q_a}) $. 

In our case, starting with $Q_a$, obeying the $SU(2)$-algebra
(\ref{6}), we introduce the
$SU(2)\times SU(2)$-algebra (\ref{6}), (\ref{9}) of $Q_a, ~K_a$. 
The set $Q_a, ~K_a$ can be considered as the analogue of the
phase space.  The analogue of the polarization condition implies
that we must choose an irreducible representation of $Q_a, K_a$. 
As we have seen already, the smooth manifold limit corresponds to
$q,k\rightarrow \infty$.   For a given irreducible
representation, labelled by the spin values $(q,k)$, there are
several representations possible for the angular momentum
$J_a=\lambda (Q_a +K_a)$, the lowest possible $j$-value being
$\vert q-k\vert$.  The difference $\vert q-k\vert$ may be
interpreted as the strength of a magnetic monopole at the center
of the sphere, or a uniform magnetic field through the sphere. 
(I thank Polychronakos for discussions clarifying this point.) 
In the absence of any magnetic monopole field, we can take $q=k$.
\vskip .1in
\noindent{$\underline{{\rm Generalization ~to} ~SU(n)}$

More general brane solutions require the consideration of
$N$-dimensional representations of $SU(n)$,
$n>2$, with $N\rightarrow \infty$ eventually. The generalization
of our considerations to $SU(n)$ is straightforward. Basically
one has to consider an
$SU(n)\times SU(n)$-algebra 
\beqar &&[Q_a, Q_b]~= {i\over \l } f_{abc} Q_c\nonumber\\ &&[K_a,
K_b]~= {i\over \l} f_{abc} K_c\nonumber\\ &&[K_a,
Q_b]~=0\label{22}
\eeqar The momentum operator can then be defined by
\beq P_a = { n\l \over 2} ~f_{abc} {K_b Q_c \over \sqrt{Q^2K^2}}
\label{23}
\eeq
$P_a$ is a derived quantity, with $Q_a,~K_a$ defining the basic
algebra, as in the case of $SU(2)$.  The commutator of $P_a$ with
$Q_b$ can be evaluated without too much trouble, eventhough it is
more involved than in the case of
$SU(2)$. The following identity for the the structure constants
is useful for this calculation. Let
$t^a$ be hermitian $(n\times n)$-matrices which form a basis of
the  Lie algebra of $SU(n)$ with $[t^a,t^b]=if^{abc}t^c,~\Tr
(t^at^b)= 
\half \delta^{ab}$. We can then write
\beqar f_{amc}f_{bkc} +f_{bmc}f_{akc} &&= {\partial^4 \over
\partial x^a 
\partial x^b \partial y^m \partial y^k} ~F\nonumber\\ F &&=-\Tr
\left\{ [t\cdot x ,t\cdot y] ~[t\cdot x,t\cdot
y]\right\}\label{24}
\eeqar The traces can be evaluated using the identity
\beq t\cdot x ~t\cdot y +t\cdot y ~t\cdot x = {x\cdot y\over n}
+2 ~d_{abc} x_a y_b ~t_c
\label{25}
\eeq where $d_{abc} = \Tr \{ (t_a t_b+t_bt_a )t_c \}$.
Equation(\ref{24}) then becomes the identity
\beqar f_{amc}f_{bkc} +f_{bmc}f_{akc} && =\Bigg[ {4\over
n}\delta_{ab}\delta_{mk} -{2\over n}\left(
\delta_{am}\delta_{bk} +\delta_{ak}\delta_{bm}\right)\nonumber\\
&&~~~~~~~~~~+~8 ~d_{abc} d_{mkc} ~-4 ~d_{amc} d_{bkc}~ -4 ~d_{akc}
d_{bmc}\Bigg] \label{26}
\eeqar With the help of this identity, the commutator of $P_a$
with $Q_b$ is now obtained as
\beqar [P_a, Q_b] &&= {i\over \sqrt{K^2Q^2}} \left[\delta_{ab}
K\cdot Q -\half
\left( K_a Q_b +K_b Q_a \right) \right] ~+~{i\over 2\l } f_{abc}
P_c \nonumber\\ && ~~~~~~~~~~+~{in \over
\sqrt{K^2Q^2}}~K_mQ_n~\left(2~ d_{abc}d_{mnc}~ -d_{amc}d_{bnc}
-d_{bmc} d_{anc} \right) \label{27}
\eeqar The calculation of $[P_a,P_b]$ is more involved. It does
not seem to be very illuminating for our discussion.

It is also possible to develop expressions for $Q_a,~K_a$, which
are analogues of equations (\ref{13})-(\ref{18}), in terms of an
$(n^2-1)$-vector $x_a$ which parametrizes $SU(n)$. We can write
the variation of a group element $g\in SU(n)$ as
$g^{-1}~d g = it_a E_{ab} dx^b$ and $d g ~g^{-1}= it_a {\tilde
E}_{ab}dx^b$. The quantities $E_{ab}$ and ${\tilde E}_{ab}$ are
transposes of each other. For example, if we use an exponential
parametrization $g= \exp(it\cdot x)$, we can write
\beqar E_{ab}&&= \int_0^1 d\alpha~ 2~\Tr \left( t_a e^{-i\alpha
t\cdot x} t_b e^{i\alpha t\cdot x}\right)
\nonumber\\ {\tilde E}_{ab}&&= \int_0^1 d\alpha~ 2~\Tr \left( t_a
e^{i\alpha t\cdot x} t_b e^{-i\alpha t\cdot x}\right)\label{28}
\eeqar This equation shows explicitly that $E_{ab} ={\tilde
E}_{ba}$. The left and right translation generators are then
\beqar L_a &&= ~~i {\tilde E}^{-1}_{ka}{\partial \over \partial
x^k} \nonumber\\ R_a &&= -i E^{-1}_{ka} {\partial \over \partial
x^k}\label{29}
\eeqar with $L_a g = -t_a g,~ R_a g = gt_a$. These obey the Lie
algebra relations $[\xi_a ,\xi_b ]=if_{abc} \xi_c$, $\xi = L,~R$.
In terms of $V_a
\equiv (L_a +R_a)$ and $A_a\equiv (L_a -R_a) $, this becomes 
\beqar &&[V_a ,V_b ]=if_{abc} V_c\nonumber\\ &&[V_a, A_b] =
if_{abc} A_c\label{30}\\ &&[A_a, A_b]=if_{abc}V_c\nonumber
\eeqar Since $A_a$ involves the symmetric combination $
E^{-1}_{ka} +{\tilde E}^{-1}_{ka}$,  the last of these relations
is unaltered by shifting the $A_a$ by $x_a$, i.e.,
$[A_a + x_a , A_b +x_b ]= if_{abc} V_c$. Further,
$e^{i\theta\cdot V} g= e^{-it\cdot \theta} ge^{it\cdot \theta}$,
showing that $x_a$ transforms as a vector under the action of
$V_a$. The operators $Q_a$ and $K_a$ can then be defined as
\beqar &&Q_a = ~~x_a ~+~ {1\over 2 \l } (V_a +A_a)\nonumber\\
&&K_a = -x_a ~+~ {1\over 2\l}  (V_a -A_a )\label{31}
\eeqar These can be used as the starting pont for a large
$\l$-expansion around some chosen value of $x_a$.

\vskip .2in
\leftline{\bf Acknowledgments}

This work was supported in part by the National Science
Foundation grant PHY-9605216 and a PSC-CUNY-30 award. I thank
Professor Bunji Sakita for useful discussions.  I also thank
Professor N. Khuri and the Rockefeller University for hospitality
during part of this work and I. Giannakis for a critical reading
of the manuscript.

\end{document}